\documentstyle[12pt,rotate,psfig]{article}

\begin{document}

\newcommand{\be}{\begin{equation}}
\newcommand{\ee}{\end{equation}}
\newcommand{\ba}{\begin{eqnarray}}
\newcommand{\ea}{\end{eqnarray}}
\newcommand{\nl}{\nonumber \\ \nonumber \\}

\newcommand{\D}[1]{{\cal D}\left[{#1}\right] \,}

\begin{center}
{\bf \Large  Initial State of Disoriented} \\
\vspace{0.2cm}
{\bf \Large  Chiral Condensate Formation}

\vspace{0.7cm}
{\sc Tam\'as S. Bir\'o}
\\
\vspace{0.3cm}
MTA KFKI RMKI Theory Division \\
H-1525 Budapest Pf. 49 \\
\vspace{0.2cm}
and \\
\vspace{0.2cm}
Physics Department, University of Bergen \\
N-5007 Bergen, All\'egaten 55 \\


\vspace{0.2cm}
\end{center}

\vspace{0.5cm}
\centerline {\bf Abstract}
\vspace{0.5cm}

\renewcommand{\baselinestretch}{0.5}
\footnotesize 
It is very unlikely to hit at high temperatures the
state which is believed to be the starting point of DCC
formation in the quenched approximation. The chiral
symmetry is rather restored  by large amplitude Goldstone modes,
where energy is stored in the change of the chiral angle in space
(domain to domain) and time. Such iso-p-wave states carry more entropy.
In this paper we attempt to treat this contribution
in the mean field approximation.

\renewcommand{\baselinestretch}{1.0}
\normalsize
\vspace{0.7cm}
\noindent
One of the most interesting recent suggestions of exotic
phenomena in high energy heavy ion collisions is the
possibility of the formation of disoriented chiral
condensates (DCC) \cite{BKT93}-\cite{Pi95}. This idea has been built
on the semiclassical picture of chiral symmetry restoration
at high temperature where finite temperature effects were
considered on the pressure, such as masses of pion and sigma meson
to the subleading order \cite{GGP94}-\cite{AHW95}.

\vspace{0.2cm}
Recently many works on various aspects of DCC formation
has been published \cite{CMM96}-\cite{MMu95}.
Usually these considerations assume an initial state at high
temperature in which the chiral symmetry is restored by
vanishing collective fields, and independent vibrations in each
isospin direction are present as thermal fluctuations.
This configuration sits on the top of the barrier of the
potential energy at zero temperature, so a sudden cooling of
the system supposedly brings it into an unstable state.

\vspace{0.2cm}
Indeed, while at low temperature - in the slightly asymmetric
version of the model - a symmetry breaking vacuum state
with finite sigma values can be consistently obtained already in
the mean field approximation, at high temperatures no such 
meson condensate is found.
Instead of the heavy sigma and light pion masses 
somewhat heavier longitudinal and somewhat lighter transverse masses
of the fluctuating fields can be derived \cite{GMu94}.
The internal isospin direction of the meson fluctuations is, however, 
indefinite at high temperature, hence the original pion and sigma
modes have the same average mass square.
These masses furthermore approach zero as the expectation value
of the order parameter \hbox{$\langle \Phi^2 \rangle = \sigma^2 + \pi^2$}
vanishes according to an effective potential derived in this
approximation.

\vspace{0.2cm}
The point of our present criticism is that these considerations
neglect  the possibility of internal rotations in the O(4)
symmetric isospin space; the finite temperature Goldstone modes.
Exactly in energetic heavy ion collisions it is hardly
imaginable to prepare a state with sufficiently high temperature
to overcome the barrier and restore the chiral symmetry on the
one hand and not exiting any Goldstone modes (pions) on the other hand.

\vspace{0.5cm}
Let us consider first the simple Lagrangian of the linear sigma model,
\be
{\cal L} =  \frac{1}{2} \partial_{\mu}\Phi^a\partial^{\mu}\Phi_a
\, - \, V(\Phi^a\Phi_a)
\ee
with $\Phi_a=(\sigma,\vec{\pi})$ (4 components),
and
\be
V(z) = \frac{\lambda}{4} \left( z - f^2 \right)^2
\ee

self-interaction potential. Here $f=f_{\pi}$ is the pion
decay constant and $\lambda \approx 20$ is the coupling constant
of the linear sigma model.
Later a symmetry breaking term, $Hn_a\Phi^a = H_a\Phi^a$,
can be added to the Lagrangian with
\hbox{$H = f_{\pi}m_{\pi}^2$} and $n_a=(1,0,0,0)$ pointing
into the isospin direction of the sigma field.
Also the term $f^2$ will be altered in this latter case
to \hbox{$f^2 = f^2_{\pi} - m^2_{\pi}/\lambda$}.

\vspace{0.2cm}
At high temperature the condensate vanishes. The classical state
around which one may consider fluctuations is a solution of
the classical equations of motion
\be
 \Box \Phi^a +\lambda( \Phi^c \Phi^c - f^2 ) \Phi^a = H^a,
\label{CLEOM}
\ee
where \hbox{$\Box = \partial_{\mu}\cdot \partial^{\mu}$}
is the d'Alambert operator.
This is usually treated as four coupled equations,
at high temperature adding only a fluctuative thermal 
contribution to the O(4) symmetric length of the meson fields,
in the one loop approximation $T^2/2$.
This is, however, insufficient because the kinetic energy 
of the angular motion in the isospin space has a 
non vanishing thermal average, too.

\vspace{0.2cm}
The classical problem possesses symmetries and accordingly
conserved Noether currents. The above equation is to be solved
for a fixed value of the Noether charges or after averaging
thermally their contribution. It is easy to see that this cannot
vanish at high temperature.
The Noether current related to rotations in the four dimensional
isospin space is given by
\be
I_{\mu}^{ab} = \frac{1}{\sqrt{2}}
\varepsilon^{abcd} \Phi_c \partial_{\mu} \Phi_d
\ee
with \hbox{$\varepsilon^{abcd}$} being the totally antisymmetric
four dimensional tensor. This four dimensional antisymmetric tensor
can be represented by two three-vectors. They are called the
vector and axialvector current, respectively. The axialvector
current is slightly not conserved because of the explicit
symmetry breaking in the $\sigma$ direction:
\be
\partial^{\mu} I_{\mu}^{ab} =
\frac{1}{\sqrt{2}} \varepsilon^{abcd} \Phi_c H_d.
\ee
The kinetic energy of the mean field 
can be rewritten with the help of this expression as
\be
 \frac{1}{2} \partial_{\mu}\Phi^a\partial^{\mu}\Phi_a =
 \frac{1}{2} \partial_{\mu}\Phi\partial^{\mu}\Phi  + 
 \frac{I_{\mu}^{ab} \cdot I^{\mu}_{ab}}{2\Phi^2}
\ee
where \hbox{$\Phi^2 = \Phi^c \Phi_c$} is the "isospin radial"
field component in the four dimensional space. 
This kinetic energy term corresponding to
nonzero Noether currents describes angular motion in the
isospin space.

\vspace{0.2cm}
It is straightforward to see that the often investigated initial
condition of DCC formation, $\Phi^2=0$, can only be realized
if the Noether current vanishes. This state is a very special
state which carries almost no entropy compared to the
isospin angular motion ("p-wave") states. Therefore at high
enough temperature the characteristic states have finite 
\hbox{$I^2=I_{\mu}^{ab} \cdot I^{\mu}_{ab}$}
and $\Phi^2$ according to the equipartition principle
\be
\frac{I^2}{2\Phi^2} \propto T^4.
\ee
At high temperature supposedly 
only thermal fluctuations would contribute to the quadratic
expectation values (semiclassically $\Phi^2 \approx T^2$).
Taking into account the above estimate for the iso-rotational
energy, however, we arrive at
a Noether charge density $I \approx T^3$.
Indeed this can be interpreted as the physical pion density.

\vspace{0.2cm}
In order to make a first attempt to solve the problem of coupled
iso-rotation and condensate formation we assume that in the
thermal bath the total energy density is partitioned
equally between four kinetic and one potential energy term
(i.e. only the iso-radial component feels a potential, the
iso-angular motion is free).
The mean field equation for $\Phi$ can be derived by varying
the Lagrangian with respect to $\Phi$ by keeping
\hbox{$N^a=\Phi^a/\Phi$} constant. In the symmetric model ($H=0$)
it contains then an extra ''centrifugal'' term in the effective potential,
\be
\Box \langle\Phi\rangle + \left[
\lambda (\langle \Phi \rangle^2 + \frac{1}{2} T^2 - f^2 )
- \frac{\gamma T^4}{\langle \Phi \rangle^2}
\right]\langle\Phi\rangle = 0.
\ee
Here \hbox{$\gamma = 4\pi^2/75 \approx 0.47$} 
is estimated for the partition of kinetic energy of
the rotational motion in the four dimensional 
isospin space (3/5 of total thermal energy).
The brackets \hbox{$\langle \,\, \rangle$} denote
an average over the angular motion. The thermal average of
vibrational fluctuations (waves) is taken into
account already in the self-energy term,
\hbox{$T^2/2$}, to leading order.

\vspace{0.2cm}
This mean field equation has a static solution at each
temperature which can be analytically obtained solving
the corresponding second order algebraic equation.
In Fig.1 the scaled order parameter 
\hbox{$\langle  \Phi \rangle^2 / f^2$}
is shown as a function of the scaled temperature
$T/T_c$ for different values of the coupling constant $\lambda$. 
The critical temperature obtained in the
nonlinear sigma model \hbox{($1/\lambda = 0$)} is
\hbox{$T_c = f_{\pi} \sqrt{2} \approx 131$ MeV}.
This simple estimate shows the physical point: at the
realistic value of $1/\lambda = 0.05$ the
$\langle \Phi^2 \rangle= 0$ state is already
unreachable. Moreover according to the perturbative
philosophy of the linear sigma model $\lambda$ has to be small.
The uppermost curve on Fig.1, which corresponds to 
\hbox{$\lambda = 0.3$}, reaches the outer wall of the mexican
hat potential (\hbox{$\langle \Phi \rangle \ge f$} values)
already at \hbox{$1.3T_c \approx 170$ MeV} 
and never comes even close to zero.

\begin{figure}[h]

\centerline{\psfig{figure=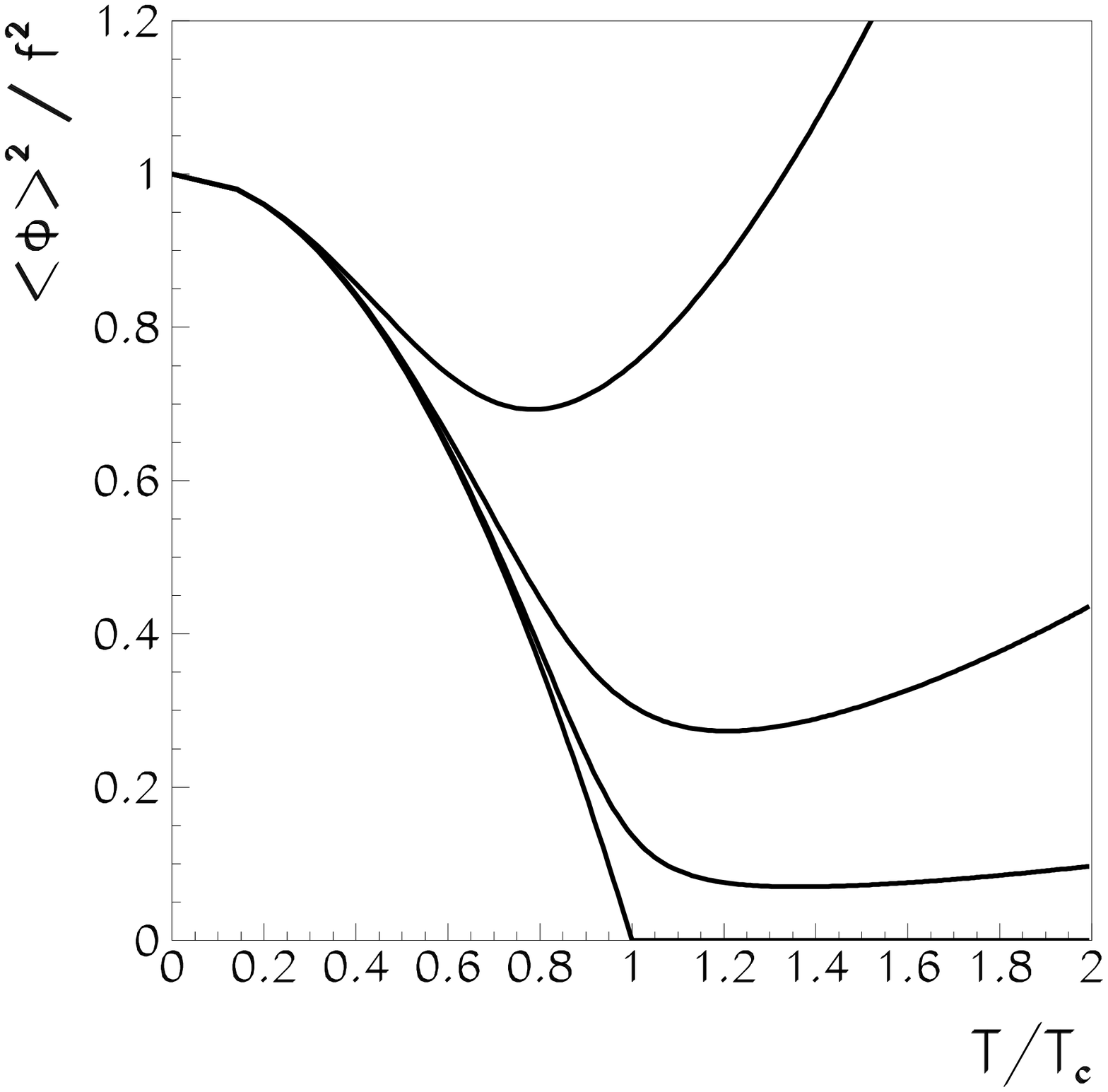,height=12.0cm,width=9.0cm}}
\vspace{0.3cm}
\caption[ROT]{The quadratic strength of collective meson fields, 
	\hbox{$\langle\Phi\rangle^2/f^2$}, is plotted against the
	scaled temperature $T/T_c$. The upper curves belonging to
	\hbox{$\lambda=0.3, \, 20$} and \hbox{$100$} 
	respectively show the ''centrifugal'' effect due to the 
	angular motion in isospin space, 
	while the lowest one the conventional 
	picture obtained for \hbox{$1/\lambda=0$}. }

\end{figure}

\vspace{0.2cm}
Although in our opinion it is highly unlikely to prepare a
state with $I=0$ at high temperature, it might not be impossible
to produce disoriented chiral condensates if the cooling process
is fast enough. The system should expand so fast, that the
isospin angular motion freezes at the "wrong" chiral angle.
In this case its kinetic energy would be converted into fluctuations
localized around the false chiral vacuum and only on a slower
timescale would the sigma vacuum appear.
The energy contained initially also in high $I$ collective
fields could so be converted into many pionic fluctuations
(small amplitude iso-vibrations) around the symmetry breaking
sigma vacuum at $I=0$. The details of such a process are,
however, at the moment unclear to us.

\vspace{0.2cm}
In the perturbative scenario ($\lambda = 0$) it is never possible
to sit on the top of the barrier and there are many excited
Goldstone modes at high temperature. 
In a scenario close to the nonlinear sigma
model ($1/\lambda \approx 0.01$) and near to the critical temperature 
\hbox{$T_c=f_{\pi}\sqrt{2}\approx 130$ MeV} quite low values
of $\langle \Phi \rangle$ may be realized before the typical
motion emerges again to the outer wall of the mexican hat potential
with the further rise of the temperature \hbox{$(T > 2T_c)$}.
Only in the case $1/\lambda = 0$ reaches the collective component
zero, but one should not forget that the sum with the average
of the thermal fluctuations, i.e. 
\hbox{$\langle \Phi \rangle^2 + T^2/2 = f^2$}
is restricted to be a constant in the nonlinear sigma model.

\vspace{0.7cm}
In conclusion a \hbox{$\langle \Phi \rangle^2 \approx 0$}
state at high temperature is rather unrealistic.
Consequently the traditional starting point of the
disoriented chiral condensate formation in the quenched
approximation is misleading and more realistic calculations
have to be carried out in order to judge the chances
of DCC formation in relativistic heavy ion collisions.

\vspace{0.2cm}
In the future this problem can be attacked
in the following way:
In principle the internal rotational motion in the
O(4) isospin space can be separated from the radial motion.
At finite temperature instead of the partition function of
vibrations in each Fourier mode, these rotations have to be
summed in an effective Lagrangian. Unfortunately even free 
rotation cannot be summed in an analytical expression like the
Bose distribution stemming from a vibrational spectrum.
Instead two possible approaches are available: i) carry out
semiclassical simulations using the field equation (\ref{CLEOM})
with an initial condition corresponding to a finite $I$ 
or ii) integrate out the isospin angular motion from the partition
function by using auxiliary fields in the functional integral 
by resolving the radial constraint similar as
Bochkarev and Kapusta have done\cite{BKa95,BKa96}. We deal with the
second possibility in a forthcoming paper. 


\vspace{0.7cm}
{\bf Acknowledgment}
I acknowledge the financial support and the warm
hospitality of the University of Bergen.
Enlightening discussions with L.P. Csernai, A.A. Adrianov and
\'A. M\'ocsy are gratefully acknowledged.
This work has also been supported by the hungarian
National Science Research Fund (OTKA T014213).

\nopagebreak[9]


\end{document}